# Intense white luminescence in ZnTe embedded porous silicon

O. de Melo,[1] C. de Melo,[1] G. Santana,[2] J. Santoyo,[3] O. Zelaya-Angel,[3] J. G. Mendoza-Álvarez,[3] and V. Torres-Costa[4]

[1]*Physics Faculty, University of Havana, La Habana, Cuba*
[2]*Instituto de Investigaciones en Materiales, Universidad Nacional Autónoma de México, Cd. Universtaria, A.P. 70-360, Coyoacán 04510, Mexico*
[3]*Department of Physics, CINVESTAV, IPN, Mexico*
[4]*Departmento de Física Aplicada, Universidad Autónoma de Madrid, Cantoblanco, 28049 Madrid, Spain*

Porous silicon layers were embedded with ZnTe using the isothermal close space sublimation technique. The presence of ZnTe was demonstrated using cross-sectional energy dispersive spectroscopy maps. ZnTe embedded samples present intense room temperature photoluminescence along the whole visible range. We ascribe this PL to ZnTe nanocrystals of different sizes grown on the internal pore surface. Such crystals, with different orientations and sizes, were observed in transmission electron microscopy images, while transmission electron diffraction images of the same regions reveal ZnTe characteristic patterns. © *2012 American Institute of Physics*. [http://dx.doi.org/10.1063/1.4731276]

Luminescence of porous silicon (PS) has been a subject of strong research in the last two decades.[1] Although many points regarding the origin of the luminescence are still not clear, there is a consensus in attributing it to quantum confinement (QC) in silicon nano-rods.[2,3] In fact, theoretical models predict that Si nanocrystals with adequate sizes can be responsible for emission in the whole visible range. However, oxygen passivation of the Si internal surface may avoid short wavelength emissions from nanocrystals smaller than around 3 nm, and hence red emission is frequently observed in most reports.[1] In fact, blue or green emission has also been reported in oxygen free samples.[4]

The infiltration of PS with different materials has allowed modifying the properties of either the PS matrix or the embedded material. For example, nanocrystalline ZnO films were synthesized on porous PS layers using radio frequency sputtering; intense blue luminescence was observed along with the standard red emission.[5] Atyaoui *et al.*[6] presented results for Ce doping effects on photoluminescence (PL) properties of PS. In particular, white luminescence devices are an active research topic because their applications in the technologies of full-color displays[7] or lighting.[8]

In this paper, isothermal close space sublimation (ICSS) has been used to embed ZnTe in PS using cyclic exposures of the PS surface to Zn and Te sources. Infiltration of the pores is demonstrated by cross-sectional energy dispersive spectroscopy (EDS) compositional analysis. White room temperature (RT) PL was observed in the range of 400–700 nm for samples embedded at different conditions. This luminescence could be observed at the naked eye as a white spot under a HeCd laser excitation.

PS layers were formed by electrochemically etching $1 \times 1$ cm crystalline Si $\langle 100 \rangle$ wafers (p-type, 0.05–0.1 $\Omega$ cm) using an HF-based solution and applying a current density of 150 mA/cm$^2$ for 20 s, conditions known to produce nanoporous sponge-like PS.[9] The etched area was a circle of 8 mm in diameter. ZnTe semiconductor was grown by the ICSS technique. General details of the experimental setup and process of this technique can be found elsewhere.[10] A graphite sample holder is alternately exposed to Zn (99.99%) and Te (99.999%) elemental sources (provided by GOODFEL-LOWS) at 385 °C with a purge step in between. The process is repeated a given number of cycles and carried out under high vacuum of $10^{-5}$ Torr. A programmed linear actuator LA12-PLC from LINAK was used to perform the deposition cycles. Previously to the growth experiments, PS substrates were degreased in acetone and isopropyl alcohol. In some growth experiments, the samples were subjected to a further HF:H$_2$O (1:10) etching for 15 s. Exposure times to the elements and purge times were varied. Due to the geometry of the graphite sample holder, the area exposed to Zn and Te was a circle but with 0.7 mm in diameter, concentric with (but smaller than) the PS circular surface. Thus, 1 mm outer ring of the PS surface was not exposed to Zn and Te during the growth process since this region was in close contact with the graphite substrate holder and covered by it. As a consequence, characterization of ZnTe embedded PS and ZnTe-free PS was possible on the same sample. The optical characterization of both zones, PS region and ZnTe embedded PS region, was carried out by using a PL system with the 325 nm wavelength He-Cd laser line (maximum output power of 16 mW), as excitation source. To determine the structure of the samples, grazing incidence x-ray diffraction (XRD) scans were performed using a Siemens D-5000 powder diffractometer with CuK$_{\alpha 1}$ radiation. The Zn and Te were identified inside of PS thin film by mapping in a QUANTAX EDS Bruker 127 eV detector, in a field emission gun-scanning electron microscope (FEG-SEM), Carl Zeiss Auriga at 20 kV. ZnTe nanoparticles were observed in high resolution transmission electron microscopy (TEM) with their Selected Area Electron Diffraction (SAED) pattern respective in a TEM-JEM-2010, Jeol at 200 kV and 110 $\mu$A. Indexation of SAED patterns was performed taking a camera distance $L$ of 20 cm and wavelength $\lambda$ of 0.0027 nm.

Fig. 1 shows an x-ray diffractogram of a PS sample embedded with ZnTe at grazing incidence. As can be seen, only peaks corresponding to the stable ZnTe zinc blend structure are observed (the peak at around $2\theta = 56°$ was identified as coming from the background or the (311) Si plane).

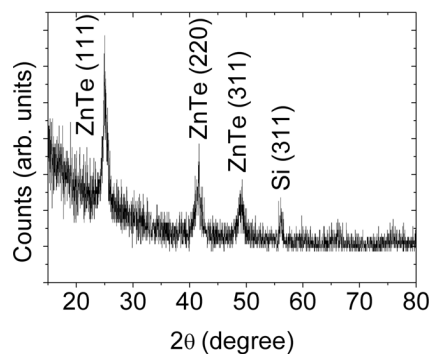

FIG. 1. Grazing incidence diffractogram for a ZnTe embedded PS layer.

Fig. 2 presents the cross sectional energy dispersive spectroscopy analysis. In Fig. 2(a) a cross sectional SEM image taken from a cleaved sample is shown; the frame indicates the regions in which EDS maps were taken. Figs. 2(b) and 2(c) show such maps at L$\alpha$ for Te and K$\alpha$ for Zn, respectively. From these images, it can be concluded that infiltration of ZnTe was achieved and that XRD peaks observed in Fig. 1 come from ZnTe embedded in the PS matrix. Energy dispersive x-ray (EDX) analysis indicates also the presence of Si and Oxygen and similar maps of that shown in Fig. 2 were taken also for these elements. These maps were taken for a sample grown with 10 cycles using exposure times to Zn and Te of 90 s and purges times after exposures of 90 s as well. However, infiltration of the pores was demonstrated also in samples grown with much smaller exposure (10 s) and purge times (5 s). A detailed study of the influence of the growth parameters on the infiltration of the pores will be presented in a separated paper.

Fig. 3 (lower spectra) represents normalized RT PL for the two different regions (represented schematically in the inset). As it can be seen in the spectra, the border of the porous silicon that was not embedded with ZnTe shows the typical red luminescence of these PS samples. In the central region, where ZnTe infiltration occurred, a wide band luminescence is observed from 400 to 700 nm, i.e., in the whole visible range. Oscillations are due to interference effects as it is demonstrated by comparing the maxima and minima of the PL with the reflectance spectrum of the same sample shown in the upper part of the figure. Both red and white bands were observed at the naked eye when the sample was illuminated with the HeCd laser. The red band is not more observed in the central region; instead, a low intensity IR band is detected. This band is in a different wavelength range than that of the pristine PS; it could be probably assigned to a defect band of ZnTe.

Luminescence in the whole visible range has been observed in carbon-rich silicon oxide obtained by carbonizing PS.[11] Since our growth experiments were made using a graphite crucible, the presence of carbon would be justified. The region more exposed to the graphite crucible was the border of the sample which was in contact with the sample holder. However, as stated above, this region does not present white luminescence. Moreover, carbon was insignificant in the EDS maps. Also ZnO has been observed to produce bright white luminescence when embedded in PS using a sol gel process.[12] The formation of ZnO in our samples could have been favoured by the exposure to Zn of the oxidized internal surfaces of PS. However, the diffractogram of Fig. 1 excludes the presence of a significant amount of ZnO in the ZnTe embedded samples. Then, we conclude that ZnTe in itself is the responsible for the observed white luminescence.

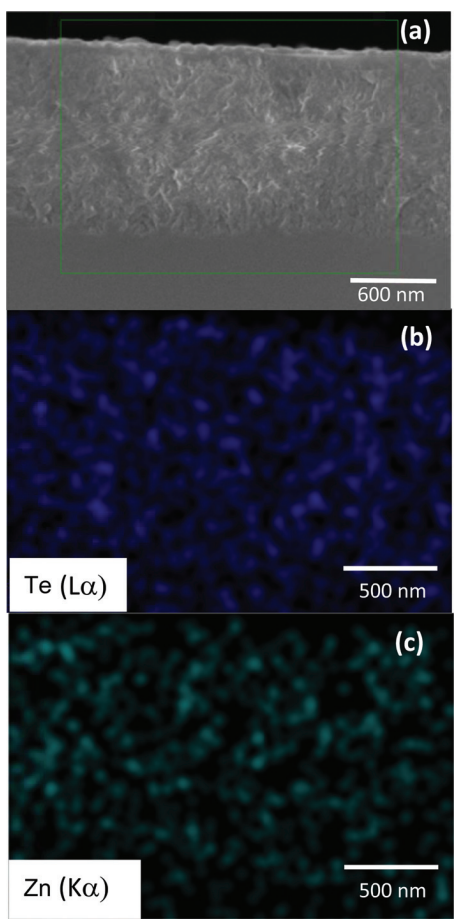

FIG. 2. (a) Cross section image of a ZnTe embedded sample. The frame indicates the region in which EDS maps were taken. (b) and (c) EDS maps for Te and Zn elements, respectively.

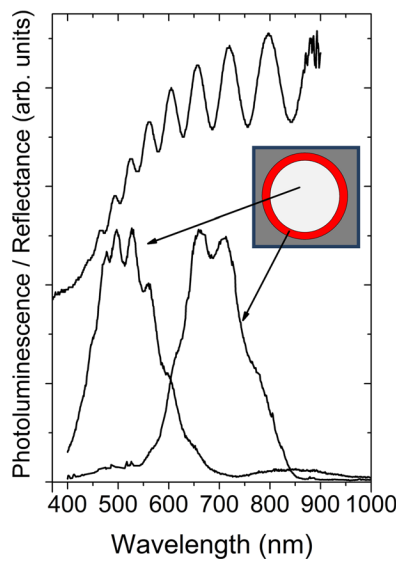

FIG. 3. Normalized room temperature photoluminescence (lower part) for the two different regions indicated in the inset: the central and outer regions. In the upper part, the reflectance spectrum of the same sample is shown.

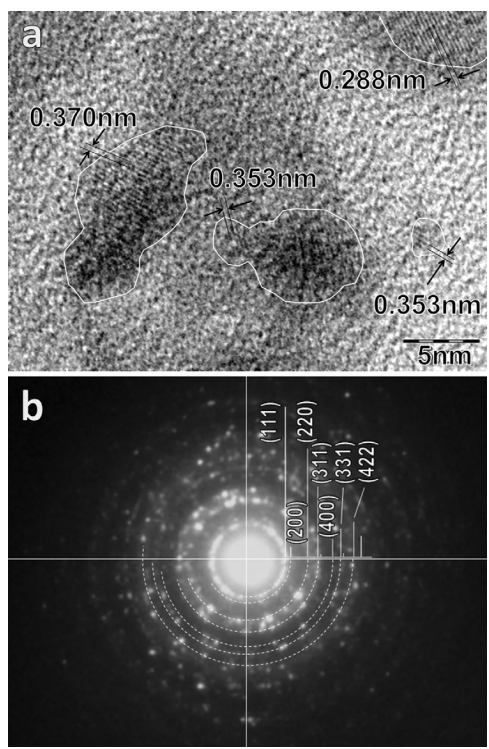

FIG. 4. (a) HRTEM image showing a region with ZnTe nanocrystals of different sizes and orientations. (b) SAED pattern of the same region. A reference diffractogram of ZnTe is superimposed showing the indexation of the circles corresponding to the different crystalline planes.

Bulk ZnTe has a direct band-gap of 2.26 eV (around 529 nm) at room temperature and a Bohr exciton radius of 6.2 nm and then it cannot emit light above this band gap value. So, intense emission observed over all visible range (even at the naked eye on the HeCd laser excitation regions) can be supposed to be promoted by ZnTe nanocrystals with different sizes, embedded in the pores of the PS layers. In fact, quantum confinement effect of the electron-hole pairs generated by the light absorption can result in an increased band gap of nanocrystals, leading to a blue shift (dependent of the crystal size) in the PL emission.[13] Due to the relatively high energy position of the valence band of ZnTe,[14] mainly type II quantum dots (QD) have been reported so far. For example, ZnTe caped with ZnSe have shown emission from blue to amber (in dependence of a thickness of ZnSe cap layer) due to a type II band alignment.[15] Large blue-shifts of the PL have been observed for ZnTe QD capped with ZnS obtained by metalorganic chemical vapor deposition technique.[16]

To check for this assumption, TEM measurements were carried out in scraped powders from a ZnTe embedded PS sample. TEM image in Fig. 4(a) shows crystallites of different (nanometric) sizes surrounded by an amorphous environment. Crystal planes in this image are identified as belonging to the {200} or {111} family of ZnTe crystals with different orientations as can be expected if ZnTe nanocrystals indeed grow on the amorphous internal surface of PS. In Fig. 4(b) the rings of SAED pattern shows the diffracted planes of ZnTe structure in the same displayed region. The distance between the rings corresponds to the interatomic distances $d$ of crystalline ZnTe. A reference diffractogram of ZnTe is superimposed to enable the identification of the diffraction spot circles as corresponding to crystalline planes of ZnTe. As can be observed, this is a characteristic pattern of polycrystalline ZnTe. According to these observations, we can suppose that ZnTe nanocrystals with different sizes grew onto the amorphous $SiO_2$ inner surface of PS. These images seem to support the assumption of ZnTe nanocrystals being responsible for the white luminescence of the samples. Due to the very large band gap of $SiO_2$, a type I band alignment can be plausible in ZnTe/$SiO_2$ heterostructures with a very large confinement potential. However, the low energy tail of the white band above around 550 nm (the band gap of ZnTe) cannot be explained without considering another PL mechanism. Other transitions due, for example, to surface states, defects, or the residual emission from PS can be responsible.

In summary, it was demonstrated that the simple physical transport ICSS technique can be used for the infiltration of sponge like structure of PS layers around 1 $\mu$m in thickness. ZnTe embedded samples showed an intense white PL which was not present in the pristine PS layer. We ascribe this luminescence to quantum confinement in ZnTe nanocrystals grown on the internal $SiO_2$ surfaces of the pores. TEM images and SAED patterns seems to support this assumption. ICSS infiltration of PS represents a simple and inexpensive route to obtain white luminescence in PS.

O.dM and O.Z.A. wish to thank the support of Instituto de Ciencia y Tecnología del DF (ICyTDF).